\documentclass[12pt]{article}
\usepackage{amssymb}
\usepackage{euscript}

\usepackage[latin1]{inputenc}
\usepackage[T1]{fontenc}
\usepackage{amsmath}
\usepackage{amsfonts}
\usepackage{amssymb}
\usepackage{latexsym}
\usepackage[english]{babel}
\usepackage[dvips]{graphicx,color}
\usepackage{latexsym}
\usepackage{graphicx}
\usepackage{graphics}
\usepackage{pstricks}

\begin{document}

\newcommand{\be}{\begin{equation}}
\newcommand{\ee}{\end{equation}}
\newcommand{\epe}{\end{equation}}
\newcommand{\bea}{\begin{eqnarray}}
\newcommand{\eea}{\end{eqnarray}}
\newcommand{\ba}{\begin{eqnarray*}}
\newcommand{\ea}{\end{eqnarray*}}
\newcommand{\epa}{\end{eqnarray*}}
\newcommand{\ar}{\rightarrow}

\def\A{\tilde{a}}
\def\G{\Gamma}
\def\r{\rho}
\def\D{\Delta}
\def\R{I\!\!R}
\def\l{\lambda}
\def\D{\Delta}
\def\d{\delta}
\def\T{\tilde{T}}
\def\k{\kappa}
\def\t{\tau}
\def\f{\phi}
\def\p{\psi}
\def\z{\zeta}
\def\ep{\epsilon}
\def\hx{\widehat{\xi}}
\def\a{\alpha}
\def\b{\beta}
\def\O{\Omega}
\def\M{\cal M}
\def\g{\hat g}
\newcommand{\dslash}{\partial\!\!\!/}
\newcommand{\aslash}{a\!\!\!/}
\newcommand{\eslash}{e\!\!\!/}
\newcommand{\bslash}{b\!\!\!/}
\newcommand{\vslash}{v\!\!\!/}
\newcommand{\rslash}{r\!\!\!/}
\newcommand{\cslash}{c\!\!\!/}
\newcommand{\fslash}{f\!\!\!/}
\newcommand{\Dslash}{D\!\!\!\!/}
\newcommand{\Aslash}{{\cal A}\!\!\!\!/}


\vspace{3mm}

\begin{center}

\vspace{3mm}

{\large Quantum states of the spacetime, and formation of black holes in AdS}

\vspace{.3in}
Marcelo Botta Cantcheff $^{\dag\,\ddag}$\footnote{e-mail:
bottac@cern.ch, botta@fisica.unlp.edu.ar}

\vspace{.2 in}

$^{\dag}${\it IFLP-CONICET CC 67, 1900,  La Plata, Buenos Aires, Argentina}\\

$^{\ddag}${\it CERN, Theory Division, 1211 Geneva 23, Switzerland}
\vspace{.4in}

\emph{\small{Essay written for the Gravity Research Foundation 2012 Awards for Essays on Gravitation}}

 March 28, 2012

\end{center}

\begin{abstract}
\noindent

 We argue that a non-perturbative description of quantum gravity should involve two (non-interacting) copies of a dual field theory on the boundary, and describe the states of the spacetimes accordingly. So, for instance, a complete description of the asymptotically Anti-de-Sitter spacetimes is given by two copies of the conformal field theory associated to the global AdS spacetime.  We also argue that, in this context, gravitational collapse and formation of a black hole may be described by unitary evolution of the dual non-perturbative degrees of freedom.

\end{abstract}

\vspace{0.7cm}

\textbf{Introduction}

\vspace{0.5cm}

The AdS/CFT correspondence represents the paradigmatic case of gravity/gauge duality where spacetime may be defined
 as emergent from a gauge theory on its conformal boundary \cite{adscft}.
 An exact description of quantum gravity is expected to be built up in general, based
 on this and other examples of holography; however, we do not have yet a deep description
  as how the states of the spacetime and gravitational dynamics emerge from the
   degrees of freedom of quantum field theory \cite{llm}.
 Furthermore, although the unitary evolution of gravitational systems is a foundational property of this holographic approach, the problem of describing the gravitational collapse into black holes with no information lost has not been completely clarified \cite{infoBH1,infoBH2,infoBH3,infoBH4}, even in conventional AdS/CFT \cite{bala-info}. In this familiar case, quantum gravity\footnote{Probably formulated as a string theory} (QG) with fixed $AdS_{5} \times S^5$ asymptotic spacetime behavior, is equivalent to a CFT on its conformal boundary $B_4 = S^3 \times {\mathbb R}$. We may view this CFT as the non- perturbative description of QG. The standard interpretation is that the exact bulk geometry $M_0 \equiv AdS_{5} \times S^5$, corresponds to the fundamental state $\left|0\right\rangle$ of the CFT Hilbert space ${\cal H}$, and general classical spacetimes should be encoded in non trivial choices of that state.
The main issue of the holographic formulation of quantum gravity is precisely how to describe more general spacetimes with fixed asymptotic behavior
in terms of states in a quantum field theory. In this work, we will try to shed light on how these states should be,
in order to describe gravitational collapse with no information loss, and such that the relevant information on a classical spacetime may be extracted from them.

Let us now consider the CFT on $B_{d} = S^{d-1}\times {\mathbb R}$, and a second (decoupled) copy denoted in what follows by a tilde.
It has been argued that asymptotically $AdS_{d+1}$ spacetime with a eternal black hole corresponds to the (entangled) state \cite{eternal}:
\be \left.\left|0(\beta)\right\rangle \! \right\rangle = \sum_n \, \frac{e^{-\frac{\beta}{2} \, E_n}}{Z^{1/2}}\left|n\right\rangle \otimes \left|\tilde{n}\right\rangle
~\in{\cal H}\otimes\widetilde{{\cal H}}~~,~~\b\equiv(k_B T)^{-1}\label{BHstate}, \ee
where $\left|n\right\rangle$ are a complete basis of eigenstates of the CFT Hamiltonian $H$, and $E_n$ are the eigenvalues. This describes a thermal state of the CFT system at temperature $T$ in the \emph{thermofield dynamics} (TFD) formalism \cite{tu}\cite{ume1,ume2}.
Note, in addition, that the state (\ref{BHstate}) is invariant under the action of $\widehat{H} = H - \widetilde{H}$.

On the other hand, it has been recently observed that the connectivity of the classical spacetime is intimately related to a quantum entanglement of the degrees of freedom of the non-perturbative equivalent field theory \cite{VR}. For example, the state (\ref{BHstate}) describes a classically connected spacetime with two asymptotically AdS regions causally separated by horizons \cite{galloway} (Fig 1(b)), which according to (\ref{BHstate}) represents a \emph{quantum superposition of disconnected (asymptotically) AdS spacetimes} $\left|n\right\rangle \otimes \left|\tilde{n}\right\rangle$.
 In fact, each of the product states $\left|n\right\rangle \otimes \left|\tilde{n}\right\rangle$ should be interpreted on the gravity side as a spacetime with two disconnected components, since they clearly describe states of a system of two completely separated physical subsystems with no entanglement between them: two (non-interacting) identical copies of $CFT$, each one dual to an asymptotically $AdS$ spacetime (see Fig 1(a)).

In this essay, we will propose an extension of the AdS/CFT conjecture which manifestly captures this fact in a general statement, and allows to describe more general emergent classical spacetimes (with fixed asymptotics) from a non-perturbative dual description \cite{collapse}.
 The main consequence of this is that the process of formation of an AdS-black hole (the state (\ref{BHstate}))
  may be described by an unitary evolution operator,
which provides a solution to the information loss paradox.

The resulting model of collapse is non-conventional in many senses: in particular, it involves topology changes of the classical geometry; and then, the semiclassical evolution from initial data cannot be continuously described by GR solutions. The dual interpretation of collapsing classical spacetimes in the present approach will be not discussed here.

\vspace{0.7cm}

\textbf{Gravity/gauge$^2$ duality and quantum states of the spacetime}

\vspace{0.5cm}

The problem of giving a non-perturbative description of the collapse reduces to clarify how a entangled (thermal) state (\ref{BHstate}) may be reached from a disentangled one $\left|\psi\right\rangle \in {\cal H}$ through some unitary operator. We will suggest that the simplest solution to this problem is to consider an extension of the non perturbative dual description from the beginning:
 a direct product of two identical copies of QFT  \cite{collapse}.

There is a natural extension of general quantum field theories, built up in order to describe the thermal effects and statistical properties of the system as entanglement of its degrees of freedom with a non-interacting identical copy: the TFD formalism. According to this, one get the
the direct product of both theories (schematically $QFT^2 \sim QFT \otimes \widetilde{QFT}$), whose time evolution is generated by a decoupled hamiltonian operator $\hat{H} \equiv H-\tilde{H}$ and the operators of $\widetilde{QFT}$ are constructed from the $QFT$ ones by the \emph{tilde
conjugation rules}, or simply TFD rules \cite{kha5}.
The total Hilbert space is the tensor product of the
two state spaces $ {\cal H}^{2} \equiv {\cal H}\otimes \widetilde{{\cal H}}$.
The states in this doubled space $\left.\left|\psi\right\rangle \! \right\rangle$ include all the \emph{statistical} information of the system QFT on behalf of
  density matrices; this allows us to describe equivalently statistical ensembles of a quantum system as \emph{pure states} in a extended quantum field theory. In fact, by tracing  $ \left.\left|\psi\right\rangle \! \right\rangle \left\langle\!\left\langle \psi\right|\right.$ over one of these spaces, one recovers the conventional density operator $\rho$ on the remaining one.

Note, according to the interpretation explained above, that \emph{all the states} of ${\cal H}^{2}$ that correspond to classical spacetimes, have two conformal boundaries.
In particular a generally entangled state as (\ref{BHstate}) represents the case of classically connected spacetimes \cite{VR} with fixed local asymptotic structure, while separate states as $\left|\psi \right\rangle \otimes |\,\widetilde{\psi}\,\rangle$ represent a spacetime with two disconnected components with the same asymptotics.

So, we can propose an extension of the gravity/gauge correspondence such that \emph{the space of states of non-perturbative quantum gravity (with fixed asymptothics) is precisely given by $ {\cal H}^{2} \equiv {\cal H}\otimes \widetilde{{\cal H}}$}.

Then, a general (quantum) spacetime $M$ with fixed local asymptotics, may be expressed as a state
\be \left.\left|M\right\rangle \! \right\rangle = \sum_{n, \tilde{m}} \, G_{n\tilde{m}} \left|n\right\rangle \otimes \left|\tilde{m}\right\rangle\,
\,\in {\cal H}_{QG} \,\,\,\,\,,\,\,\,\,\,\,\,\,{\cal H}_{QG}\equiv {\cal H}^{2} = {\cal H}\otimes \widetilde{{\cal H}}
\label{geometrystate} \ee
where $G_{n\tilde{m}}$ are complex numbers, and $\left|n\right\rangle \otimes \left|\tilde{m}\right\rangle\,$ is a complete basis of $ {\cal H}^2 = {\cal H}\otimes \widetilde{{\cal H}}$. The dynamical evolution of these states is exactly generated by the hamiltonian operator of the extended field theory $\widehat{H}=H-\widetilde{H}$. We will refer to this statement as gravity/gauge$^2$ correspondence.

This may be viewed as \emph{rule} to describe general spacetimes (with two AdS asymptotic regions\footnote{The states whose conformal boundaries are connected by an Einstein spacetime, necessarily contains one horizon separating them \cite{galloway}.}) in the context of holography and has deep significance about the mechanism of emergence of the spacetime and gravity. It collects the observations of \cite{VR} since the general states of the geometry are regarded, precisely, as entangled states; and also is in agreement with the Israel's TFD description of a maximally extended black hole spacetime \cite{israel}.

Let us mention that the states $\left.\left|M\right\rangle \! \right\rangle$ that describe \emph{classical} spacetimes (with defined local structure) should be coherent in a proper sense, so as in the case of maximally extended black holes, where $G_{n\tilde{m}}=\delta_{n\tilde{m}}\,e^{-\b E_n/2}\,Z^{-1/2}$.
 We will show a simple example below where the background spacetimes are manifestly described as coherent states.

\newpage

\vspace{0.7cm}

\textbf{The Hawking-Page transition}

\vspace{0.5cm}

As an example of this picture, the Hawking-Page transition \cite{haw-page1,haw-page2} may be described in a quantum mechanical way: as a critical behavior of the quantum amplitude. In fact, the AdS-black hole spacetime is described by the state (\ref{BHstate}) in $CFT^2$; which at low temperatures reads as
\be \left.\left|0(\beta)\right\rangle \! \right\rangle =  \,\frac{e^{-\frac{\beta}{2} E_0}}{Z^{1/2}} \left|0\right\rangle \otimes \left|\tilde{0}\right\rangle + \delta(\beta) \left.\left|\xi(\beta)\right\rangle \! \right\rangle
\label{BHstate-low}, \ee
where $\xi$ is a state orthogonal to $\left|0\right\rangle \otimes \left|\tilde{0}\right\rangle$, and $|\delta(\b)|^2 \ll |Z^{-1}\,e^{-\beta E_0}|$ as $\b < E_0^{-1} \sim b$ (here $b$ is the radius of curvature of the AdS space). So at low temperatures, the probability of collapsing this state into two disconnected global AdS spaces (Fig 1(a)) is very high, compared with other geometry states.

\vspace{0.7cm}

\textbf{The gravitational collapse}

\vspace{0.5cm}

Now let us consider again the AdS/CFT standard context, i.e, spacetime geometries whose asymptotic boundaries have local structure $AdS_{d+1} \times {\cal N}$, where ${\cal N}$ denotes a compact Riemannian manifold.
Then the extended correspondence AdS/CFT$^2$, contains the eternal AdS black hole as a particular case, the coefficients $G_{n\tilde{m}}$ are given in (\ref{BHstate}).

The picture of unitary collapse that we are proposing here is based on the fact that CFT is defined on $S^{d-1}$; and then by following the standard TFD recipe \cite{ume1,ume2}, it can be shown for any ordinary quantum field theory on a finite volume space, that a pure (disentangled) state as $\left.\left|0\right\rangle \! \right\rangle \equiv \left|0\right\rangle \otimes \left|\tilde{0}\right\rangle $  is connected to (\ref{BHstate}) \emph{by an unitary operator}.
 This argument is in principle generalizable to any context where the degrees of freedom of the dual field theory are defined on a compact Riemannian manifold $\Sigma$, and the asymptotic boundary of the spacetime is $B\sim \Sigma \times {\mathbb R}$.

 The unitary operator is known as the Bogoliubov transformation $e^{iG}$. This is, then, a \emph{candidate} to describe the evolution operator dual to a gravitational collapse. Although the real dynamics of this process may be different, the \emph{existence} of this map
 suffices to avoid the information loss paradox in a non perturbative context.

Let us now consider a simple example that captures the main ingredients of the present framework.
Following Maldacena in ref. \cite{eternal}, it is useful to substitute the CFT on $S^{d-1}$ by an oscillator-like system quantized on a finite volume\footnote{This also might be interpreted as a perturbative description of fields in the gravity side, quantized on a global AdS background geometry, and whose Fock space may be canonically built with the normalizable modes \cite{giddings-gary}}, where the generator algebra of the Bogoliubov transformations is known. In these systems
the Hamiltonian operator is
\be\label{hamcan}
H= : \sum_{n,I} w_{n,I} \, a_{n}^{\:I}a_{n}^{\dagger\:I} \,:\,,
\ee
where $n$ denotes the set of numbers characterizing the discrete (positive) frequency modes, and the indices $I,J,...=1,....,f $ label the independent physical fields.
$a_n^{\dagger \:I} , a_n^I$ are conventional creation/annihilation operators, which must extended according to the TFD rules. Then we get the extended algebra:
\begin{eqnarray}
\left[a_{n}^I,a_{m}^{\dagger \:J}\right]
&=&\left[\tilde{a}^{I}_{n},\tilde{a}_{m}^{\dagger\:J}\right]
=  \delta_{n,m}\delta^{I,J},\label{alg}
\nonumber
\\
\left[a_{n}^{\dagger\:I},\tilde{a}_{m}^{J}\right]
&=&\left[a_{n}^{\dagger\:I},\tilde{a}_{m}^{\dagger\:J}\right]
=\left[a_{n}^{I},\tilde{a}_{m}^{J}\right]=
\left[a_{n}^{I},\tilde{a}_{m}^{\dagger\:J}\right]=0.
\end{eqnarray}
 The ground state in this extended theory is conventionally defined by
\be {a}^{I}_{n}\left.\left|0\right\rangle \! \right\rangle= \A_{n}^I
\left.\left|0\right\rangle \! \right\rangle = 0\,\,\,\,, \,\,\,
\left.\left|0\right\rangle \! \right\rangle \equiv \left|0\right\rangle \otimes \left|\tilde{0}\right\rangle \ee
Let us consider Bogoliubov transformations of this extended theory, such that:
\bea a_{n}^{I}(\theta_{n}) &=& e^{-iG}a_{n}^{I}e^{iG}
=\cosh(\theta_{n})a_{n}^{I} - \sinh(\theta_{n}){\widetilde
a}_{n}^{\dagger \: I}
\\
\widetilde{a}_{n}^{ I}(\theta_{n}) &=& e^{-iG}{\widetilde
a}_{n}^{ I}e^{iG}
= \cosh(\theta_{n}) a_{n}^{I} -  \sinh(\theta_{n}) {\widetilde
a}_{n}^{\dagger \: I}
\eea
As the Bogoliubov transformation is canonical, the transformed
operators obey the same commutation algebra (\ref{alg}).
 These operators annihilate the transformed vacuum
\begin{eqnarray}
a_{n}^{I}(\theta_{n})\left.\left|0(\theta)\right\rangle\! \right\rangle &=& \widetilde
{a}_{n}^{I}(\theta_{n})\left.\left|0(\theta)\right\rangle\! \right\rangle = 0,
\end{eqnarray}
\be \left.\left|0(\theta)\right\rangle\! \right\rangle =
e^{-i{G}} \left.\left|0\right\rangle \! \right\rangle  \label{tva}. \ee
Then, the Hilbert space is constructed by applying the
 creation operators $
a_{n}^{\dagger \: I}(\theta_{n}), {\widetilde
a}_{n}^{\dagger \: I}(\theta_{n})$ to $(\ref{tva})$.
 According to the proposal (\ref{geometrystate}), we shall identify this space with ${\cal H}_{QG}$, and the ground state $(\ref{tva})$ with the \emph{background} spacetime itself.

 Requiring the extra symmetry $\widetilde{\left.\left|0(\theta)\right\rangle\! \right\rangle}= \left.\left|0(\theta)\right\rangle\! \right\rangle $, and that $e^{-i{G}}$ be unitary,
the generator for these transformations is Hermitian and given by: \be G= -i
\,\,\sum_{n, I}\theta_{n}\,(a_{n}^{I} \tilde{
a}_{n}^{I}\, - \,\tilde{a}_{n}^{\dagger\:I} a_{n}^{\dagger\:I} )\,.
\ee By using a normal ordering, we obtain:
 \be \left.\left|0(\theta)\right\rangle\! \right\rangle =
e^{-i{G}} \left.\left|0\right\rangle \! \right\rangle =
\prod_{n, I}\left[\left( \frac{1}{\cosh(\theta_{n})}\right)^f
e^{\tanh(\theta_{n})\,\,\,a_{n}^{\dagger\:I} {\tilde
a}_{n}^{\dagger\:I}} \right]
\left.\left|0\right\rangle\!\right\rangle \label{tva2}. \ee
where $Z^{-1/2}= \prod_{n}\left( \frac{1}{\cosh(\theta_{n})}\right)^f$, is a well defined real number for a finite volume system.
The precise value of the parameters in a black hole is \cite{eternal}:
 \be\label{distr-equil}\tanh(\theta_{n}) = e^{-n\b/2},\ee
 in agreement with the state (\ref{BHstate}).
 This distribution of the parameters $\theta$ maximizes the thermodynamic entropy. In the TFD formalism, the thermodynamic entropy is canonically defined as the expectation value of the operator:
\begin{eqnarray}
K &=& - k_B \,\sum_{n=1} \bigg\{ a_{n}^{\dagger \: I} a^J_{n } \,\delta_{I J}\,\ln \left(
\sinh^{2}\left(\theta _{n}\right)\right) - a^I_{n} a_{n}^{\dagger \: J
} \,\delta_{I J}\, \ln \left( \cosh^{2}\left(\theta _{n}\right)\right)\bigg\}.
\label{k}
\end{eqnarray}
in the state $\left.\left|0(\theta)\right\rangle\! \right\rangle$ \footnote{This expression results from the Von-Neumann entropy $K \equiv -k_B\sum_{n=1} N_{n} \ln N_{n}$ \cite{tu}; where the number operator is canonically defined by $
N_{n}= a_{n}^{\dagger \: I}  a_{n }^I$.}. In the canonical ensemble, the state (\ref{distr-equil}) is obtained by minimizing the free energy
\begin{equation}
F= \left\langle\! \left\langle 0(\theta) \right|\right. H \left.\left|0(\theta)\right\rangle\! \right\rangle - \frac{1}{\beta } \left\langle\! \left\langle 0(\theta) \right|\right.K \left.\left|0(\theta)\right\rangle\! \right\rangle\label{f}
\end{equation}
with respect to the transformation's parameters $\theta$`s, with $\beta$ constant \cite{tu}.

Once again the states $\left.\left|0(\theta)\right\rangle\! \right\rangle$ are
eigenstates of the combination:
\begin{equation}\label{hamiltonian}
{\widehat H} = H -{\widetilde H},
\end{equation}
in such a way that ${\widehat H}$ plays the r\^{o}le of the
Hamiltonian, generating time translations in the Hilbert space of the doubled field theory.
Then, the (coherent) states (\ref{tva2}) describe stationary spacetimes and in particular, the equilibrium state (\ref{distr-equil}) corresponds to the physical (stable) situation as the eternal black hole.
 Note that the final state is determined by thermodynamic laws and it does not depend on the specific form of the field theory or its dyamics. This fact suggests that \emph{the dual of the classical gravitational collapse is the thermalization process in the non-perturbative description}. The novelty here is the generality of the proposal (\ref{geometrystate}), which implies that this association should be applicable to \emph{all} the gravitational processes\footnote{This resembles recent interpretations of gravity as entropic force \cite{verlinde}}.

Nevertheless, if we define an effective time independent Hamiltonian of interaction $H_I$, added for a while to the decoupled one, $\widehat{H}$, precisely as being a functional of the Bogoliubov generator we get a candidate to model the behavior of the fundamental degrees of freedom in a gravitational collapse, with the remarkable property that: the algebra of operators of the non perturbative field theory, so as the states of the theory, will be preserved under time evolution since $G$
  generates canonical transformations. So, let us define
\be\label{Hint}
H_I \equiv \,i \, \epsilon_T(t) \,\sum_n\, \lambda_n\,\, \G_n \,\,,
\ee
where \be \G_n = -i
\,\delta_{I J}\,\sum_{n}\,(a_{n}^{I} \tilde{
a}_{n}^{J}\, - \,\tilde{a}_{n}^{\dagger\:I} a_{n}^{\dagger\:J} )\, ;
\ee
 $\lambda_n$ are effective coupling constants which may be assumed to be small;
 and the step function $\ep_T (t) $ is defined to be $1$ in the interval $(0,T)$ and vanishing otherwise.

The free hamiltonian $\hat{H}$ commutes with $H_I$ for construction, then
in the interaction picture, the evolution from the initial state prepared as $ \left.\left|\psi(t=-\infty)\right\rangle\! \right\rangle \equiv \left.\left|0\right\rangle \! \right\rangle $, after a time $t > \tau$ results:
\be \left.\left|\psi(t)\right\rangle\! \right\rangle =
e^{-i  \int_{-\infty}^t \, H_I \,dt} \left.\left|\psi(t=-\infty)\right\rangle\! \right\rangle =
e^{ \tau\, \sum_n \lambda_n\,\Gamma_n} \left.\left|\psi(t=-\infty)\right\rangle\! \right\rangle \label{evol}. \ee
 By fixing the couplings to be $\lambda_n  \equiv \tau^{-1}\theta_{n}^{(eq)} $, where $\theta_{n}^{(eq)} $ is given by (\ref{distr-equil}), this formula agrees with (\ref{tva2}), which shows that the final state is the black hole (Eq. (\ref{BHstate})).

\begin{figure}
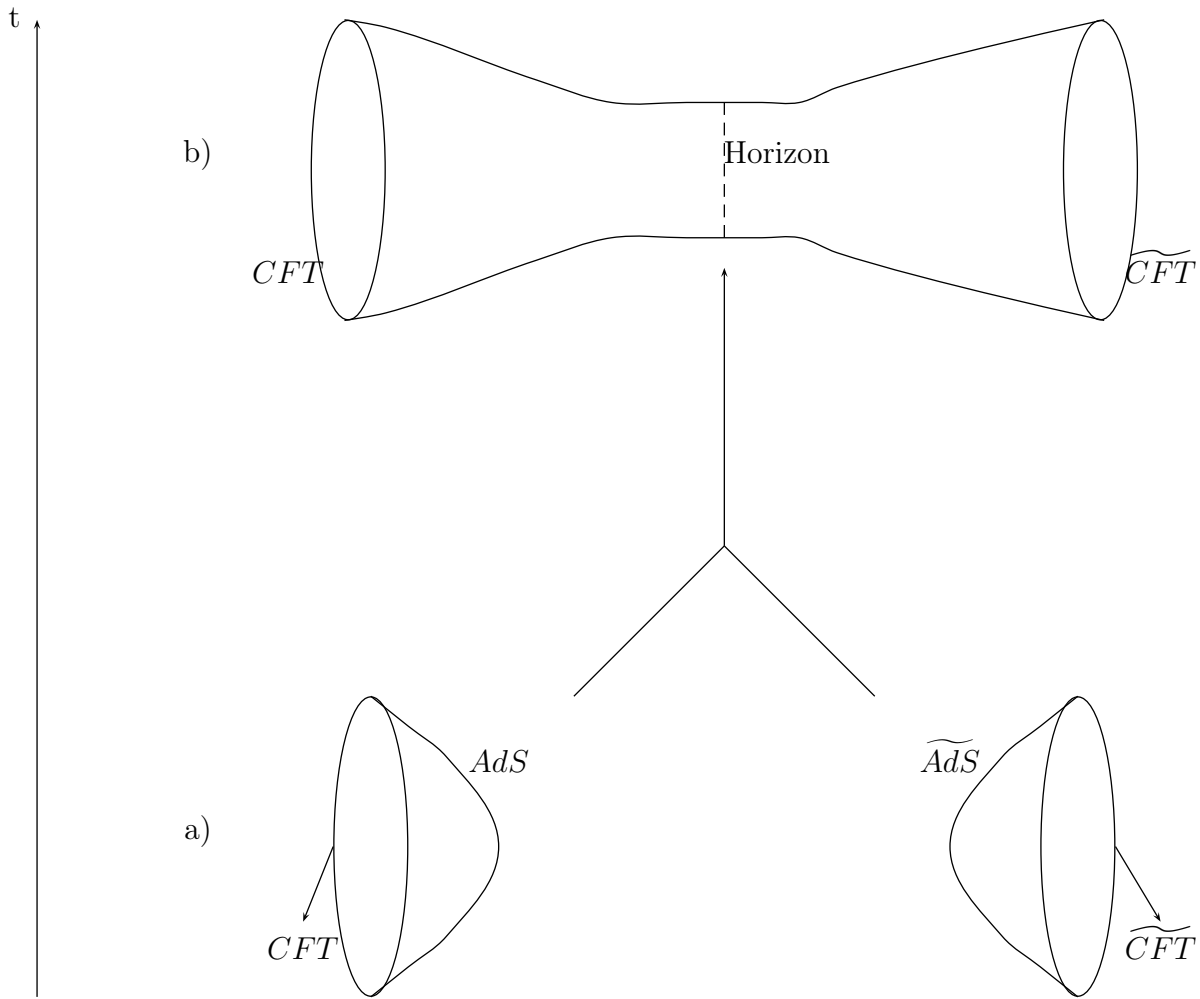

\psline[linewidth=.5pt]{->}(0,-10)(0,3)
\rput[bI](-0.3,2.9){t}
\hspace{3cm}
\psellipse[linewidth=0.5pt](1,1)(0.5,2)
\pscurve[linewidth=.5pt](0.95,3)(1.5,2.9)(3.5,2.2)(4.5,1.9)(5.5,1.9)(6,1.9)(6.5,1.9)(7,1.9)(7.5,2.1)(11.05,3)
\psellipse[linewidth=0.5pt](11,1)(0.5,2)
\pscurve[linewidth=.5pt](0.95,-1)(1.5,-0.9)(3.5,-0.2)(4.5,0.1)(5.5,0.1)(6,0.1)(6.5,0.1)(7,0.1)(7.5,-0.1)(11.05,-1)
\psline[linewidth=.5pt,linestyle=dashed](6,0.1)(6,1.9)
\rput[bI](6.7,1.1){Horizon}
\psline[linewidth=.5pt]{<-}(6,-0.3)(6,-4)
\psline[linewidth=.5pt](6,-4)(4,-6)
\psline[linewidth=.5pt](6,-4)(8,-6)
\psellipse[linewidth=0.5pt](1.3,-8)(0.5,2)
\psellipse[linewidth=0.5pt](10.7,-8)(0.5,2)
\pscurve[linewidth=0.5pt](1.3,-6)(1.8,-6.4)(2.3,-6.8)(3,-8)(2.3,-9.2)(1.8,-9.6)(1.3,-10)
\pscurve[linewidth=0.5pt](10.7,-6)(10.2,-6.4)(9.7,-6.8)(9,-8)(9.7,-9.2)(10.2,-9.6)(10.7,-10)
\psline[linewidth=.5pt]{->}(0.8,-8)(0.4,-9)
\psline[linewidth=.5pt]{->}(11.2,-8)(11.8,-9)
\rput[bI](0.2,-0.5){$CFT$}
\rput[bI](11.8,-0.5){$\widetilde{CFT}$}
\rput[bI](3,-7){$AdS$}
\rput[bI](9,-7){$\widetilde{AdS}$}
\rput[bI](-1,-8){a)}
\rput[bI](-1,1){b)}
\rput[bI](0.4,-9.5){$CFT$}
\rput[bI](11.8,-9.5){$\widetilde{CFT}$}
\vspace{10cm}
\caption{\small{(a) The state $\left|0\right\rangle \otimes \left|\tilde{0}\right\rangle$ corresponds to two disconnected global AdS spacetimes with no entanglement entropy; for other excited states $\left|n\right\rangle \otimes \left|\tilde{n}\right\rangle$ the diagram is similar but describes two asymptotically AdS parts. (b) The final state is an AdS-black hole spacetime and the entanglement entropy is a quarter of the horizon area (using the prescription \cite{takaya1,takaya2,takaya3}); the figure represents a spacial slice of the maximally extended AdS-black hole.}}
\label{figure}
\end{figure}

 In Fig.\ref{figure} we show a schematic description of the formation of a black hole according to the present model.
 This process is unitary, reversible, and is qualitatively similar to a \emph{tree diagram}, which resembles a scattering-matrix description of the quantum process of formation of a black hole \cite{smatrix}.
 Notice that both \emph{in/out} configurations correspond to stationary spacetimes and one should not attempt to interpret this in terms of any classical evolution of Cauchy data on spacial slices. In contrast, this clearly involves topology change in some intermediate stage, as expected for a quantum theory of gravity.

 \vspace{0.7cm}

\textbf{Conclusions}

\vspace{0.5cm}

In this essay, we have proposed a general form for the quantum states of the spacetime based on the Maldacena's description of AdS-black holes \cite{eternal}, and the Van Raamsdonk observations about the role played by quantum entanglement in the spacetime emergence \cite{VR}. We have argued that the formation of an AdS-black hole may be described by a unitary evolution operator in the non-perturbative dual description. It is fascinating that a kind of topology change along with the possibility of having disconnected components of the spacetime appear to be crucial ingredients of this description.
Finally, we also have pointed out the possibility of interpreting the gravitational dynamics as thermalization in the dual theory.

\vspace{12cm}

\vspace{0.7cm}

{\small \textbf{Acknowledgements} The author is grateful to L. Alvarez-Gaume, J. Russo, R. Arias, J. A. Helayel, and A. Santana. This work was partially supported by: CONICET PIP 2010-0396 and ANPCyT PICT 2007-0849.}


\begin{thebibliography}{99}

\bibitem{adscft} J.M. Maldacena,
Adv. Theor. Math. Phys. 2, 231 (1998), hep-th/9711200.


  \bibitem{llm} H. Lin, O. Lunin and J.M. Maldacena, 
JHEP 0410, 025 (2004), hep-th/0409174.



\bibitem{infoBH1}
 S. W. Hawking, \emph{Breakdown Of Predictability In Gravitational Collapse}, Phys. Rev.
D 14, 2460 (1976).
 \bibitem{infoBH2} D. N. Page, \emph{Black hole information}, hep-th/9305040.
 \bibitem{infoBH3}S. B. Giddings, \emph{Quantum mechanics of black holes}, hep-th/9412138
\bibitem{infoBH4} S. D. Mathur, \emph{The Information paradox: A pedagogical introduction}, Class. Quant.
Grav. 26, 224001 (2009)

\bibitem{bala-info} V. Balasubramanian, B. Czech, \emph{Quantitative approaches to information recovery from black holes}, Class. Quant. Grav. 28 (2011) 163001, 2011; hep-th/1102.3566

  \bibitem{eternal} J.~M.~Maldacena,
``Eternal black holes in Anti-de-Sitter,''
JHEP 0304 (2003) 021, hep-th/0106112.




\bibitem{tu}
Y. Takahashi and H. Umezawa,
Coll. Phenomena {\bf 2} (1975) 55
(Reprinted in Int. J. Mod. Phys. {\bf 10} (1996) 1755)

\bibitem{ume1} H. Umezawa, H. Matsumoto, M. Tachiki,
{\it \ Thermofield Dynamics and Condensed States}
(North-Holland, Amsterdan, 1982)

\bibitem{ume2} H. Umezawa,
{\it Advanced Field Theory: Micro, Macro and Thermal Physics }
(AIP, New York, 1993).


\bibitem{VR} M. Van Raamsdonk, Gen. Rel. Grav. 42,
2323 (2010), hep-th/1005.3035.

\bibitem{galloway} G.J. Galloway, K. Schleich, D. Witt and E. Woolgar, 
Phys. Lett. B505, 255 (2001), hep-th/9912119.

\bibitem{collapse} M Botta Cantcheff \emph{Emergent spacetime, and a model for unitary gravitational collapse in AdS}, CERN-PH-TH/2011-235, arXiv: hep-th/1110.0867



\bibitem{kha5}
A. E. Santana, A. Matos Neto, J. D. M. Vianna, F. C. Khanna,
Physica A {\bf 280} (2000) 405.




\bibitem{israel}
W. Israel, Phys. Lett. A 57, 107 (1976).








\bibitem{haw-page1} S.W. Hawking, D.N. Page, Commun. Math. Phys. 87 (1982), 577

\bibitem{haw-page2} E. Witten,
Adv. Theor. Math. Phys. 2, 505 (1998), hep-th/9803131.







\bibitem{giddings-gary}
M. Gary, S. B. Giddings, \emph{Constraints on a fine-grained AdS/CFT correspondence}, hep-th/1106.3553


\bibitem{verlinde} E. P. Verlinde, \emph{On the Origin of Gravity and the Laws of Newton}, hep-th/1001.0785.

\bibitem{smatrix} Steven B. Giddings,\emph{The gravitational S-matrix: Erice lectures}, hep-th/1105.2036














\bibitem{takaya1} S.~Ryu and T.~Takayanagi,
Phys. Rev. Lett.  {\bf 96} (2006) 181602, hep-th/0603001
\bibitem{takaya2} S.~Ryu and T.~Takayanagi,
JHEP {\bf 0608} (2006) 045, hep-th/0605073
\bibitem{takaya3} V.~E.~Hubeny, M.~Rangamani and T.~Takayanagi,
JHEP {\bf 0707} (2007) 062, hep-th/0705.0016.






\end{thebibliography}
\end{document}